\documentclass[conference]{IEEEtran}
\IEEEoverridecommandlockouts
\usepackage{cite}
\usepackage{amsmath,amssymb,amsfonts}
\usepackage{algorithmic}
\usepackage{graphicx}
\usepackage{textcomp}
\usepackage{xcolor}

\usepackage{amsthm,amsfonts,bm}
\usepackage{comment}
\usepackage{acronym}
\usepackage{cases}

\ifCLASSOPTIONcompsoc
\usepackage[caption=false,font=normalsize,labelfont=sf,textfont=sf]{subfig}
\else
\usepackage[caption=false,font=footnotesize]{subfig}
\fi

\acrodef{MIMO}[MIMO]{multiple-input multiple-output}
\acrodef{SNR}[SNR]{signal-to-noise ratio}
\acrodef{SCNR}[SCNR]{signal-to-clutter-plus-noise ratio}
\acrodef{AWGN}[AWGN]{additive white Gaussian noise}
\acrodef{KKT}[KKT]{Karush-Kuhn-Tucker}
\acrodef{SDP}[SDP]{semi-definite programming}
\acrodef{DFRC}[DFRC]{dual-functional radar-communication}
\acrodef{ULA}[ULA]{uniform linear array}

\def\BibTeX{{\rm B\kern-.05em{\sc i\kern-.025em b}\kern-.08em
    T\kern-.1667em\lower.7ex\hbox{E}\kern-.125emX}}
\begin{document}

\title{Towards a performance bound on MIMO DFRC systems
\thanks{This work was supported by the National Natural Science Foundation of China under Grants No. 62171259.}
}

\author{\IEEEauthorblockN{Ziheng Zheng}
\IEEEauthorblockA{\textit{Department of Electronic Engineering} \\
\textit{Tsinghua University}\\
Beijing, China \\
zhengzh22@mails.tsinghua.edu.cn}
\and
\IEEEauthorblockN{Xiang Liu}
\IEEEauthorblockA{\textit{Institute of Electronic Engineering} \\
\textit{China Academy of Engineering Physics}\\
Mianyang, China \\
x-liu12@foxmail.com}
\and
\IEEEauthorblockN{Tianyao Huang}
\IEEEauthorblockA{\textit{Department of Electronic Engineering} \\
\textit{Tsinghua University}\\
Beijing, China \\
huangtianyao@tsinghua.edu.cn}
\and
\IEEEauthorblockN{Yimin Liu}
\IEEEauthorblockA{\textit{Department of Electronic Engineering} \\
\textit{Tsinghua University}\\
Beijing, China \\
yiminliu@tsinghua.edu.cn}
\and
\IEEEauthorblockN{Yonina C. Eldar}
\IEEEauthorblockA{\textit{Faculty of Mathematics and Computer Science} \\
\textit{Weizmann Institute of Science}\\
Rehovot, Israel \\
yonina.eldar@weizmann.ac.il}
}

\maketitle

\begin{abstract}
It is a fundamental problem to analyse the performance bound of \ac{MIMO} \ac{DFRC} systems. To this end, we derive a performance bound on the communication function under a constraint on radar performance. To facilitate the analysis, we consider a toy example, in which there is only one down-link user with a single receive antenna and one radar target. In such a simplified case, we obtain an analytical expression for the performance bound and the corresponding waveform design strategy to achieve the bound. The results reveal a tradeoff between communication and radar performance, and a condition when the transmitted energy can be shared between these two functions.
\end{abstract}


\acresetall

\section{Introduction}

There is growing demand for implementing both radar and communication in one system \cite{DFRC1,mdy,zl,lf3} due to their similarities in both signal processing algorithm and hardware architecture \cite{Sturm,lf3}. 
To this end, \ac{DFRC} technology has attracted a lot of attention, as \ac{DFRC} design reduces system overhead and saves spectrum resources \cite{DFRC1,mdy,zl}. 
Among typical \ac{DFRC} schemes \cite{mo,lx3,zy,lx2}, \ac{MIMO}  \ac{DFRC} is of great significance because of the benefits introduced by transmit diversity  \cite{lx2,MIMOR,MIMOC}. Therefore, in this paper we analyse a fundamental performance bound of  \ac{MIMO} \ac{DFRC} systems.


Many previous works have shown that there exists performance tradeoffs between radar and communication in a \ac{DFRC} system \cite{lf3,lx}. However, the results are mostly numerical and are achieved under specific signaling and encoding schemes. Just as researchers study channel capacity as a universal performance bound for pure communication systems \cite{capacity,out,rank}, it is important to find a fundamental performance bound for a \ac{MIMO} \ac{DFRC}  system.

While it is generally difficult to simultaneously analyse the performance bounds of radar and communication, a common approach in existing works is to calculate the theoretical performance bound of one function while constraining the other  \cite{tb,lx,cl}.  In \cite{tb}, the radar performance limit is considered under a communication requirement: the transmit signal of \ac{MIMO} \ac{DFRC}  system should be carefully designed such that the down-link user receives the exact desired symbols. However, such strict signalling strategy causes radar performance degradation. Therefore, 
the \ac{MIMO} \ac{DFRC} system considered in \cite{tb} does not achieve an optimal performance tradeoff between radar and communication. 
The authors in both \cite{lx} and \cite{cl} use channel capacity to evaluate the communication performance limit, not restricted to a certain signalling scheme, under some constraint on the radar performance. However, their performance bounds are obtained by solving complex optimization problems. 

In this paper, we aim to provide an analytical expression of the fundamental performance bound of a \ac{MIMO} \ac{DFRC} system, such that the tradeoff between radar and communication functions is intuitively revealed. To facilitate the analysis, we consider a toy example in which there is only one down-link user with a single receive antenna and one radar target. Taking achievable communication rate and radar \ac{SNR} as the communication and  radar performance metrics, respectively, we formulate an optimization problem by constraining the radar \ac{SNR}  and optimizing communication performance. We obtain an analytical solution, which implies an optimal transmit design strategy that achieves the corresponding theoretical performance limit.


The remainder of this paper is organized as follows. Section~\ref{s2} introduces the system model and formulates a performance bound analysis as an optimization problem. Section~\ref{s3} gives an analytical expression of the performance bound. Simulations are performed to verify the analysis in Section~\ref{s4}. Section \ref{s5} draws conclusions.

\emph{Notation}: We use boldface lowercase letters for column vectors, boldface uppercase letters for matrices. Superscripts $\left( \cdot \right)^{H}$ and $\left( \cdot \right)^{*}$ represent Hermitian transpose and conjugate respectively, and $\mathrm{tr} \left( \cdot \right)$ stands for the trace of a matrix. The $N$-dimensional complex Euclidean space is expressed as $\mathbb{C}^{N}$. A complex Gaussian distribution with mean $\bm{\mu}$ and covariance $\bm{\Sigma}$ is expressed as $\mathcal{CN} \left( \bm{\mu},\bm{\Sigma} \right)$. The statistical expectation is represented by $\mathrm{E} \left\{ \cdot \right\}$. $\left| \cdot \right|$ and $\left\Vert \cdot \right\Vert_2$ denote absolute value and Euclidean norm, respectively.

\section{System model and problem formulation}\label{s2}

We consider a theoretical performance bound of a \ac{MIMO} \ac{DFRC} system. To this end, we first introduce the system model, the communication and radar performance metrics as well as the transmit power constraint. Then, we formulate the fundamental performance bound as an optimization problem.

\begin{figure}
	\centering
	\includegraphics[width=0.8\linewidth]{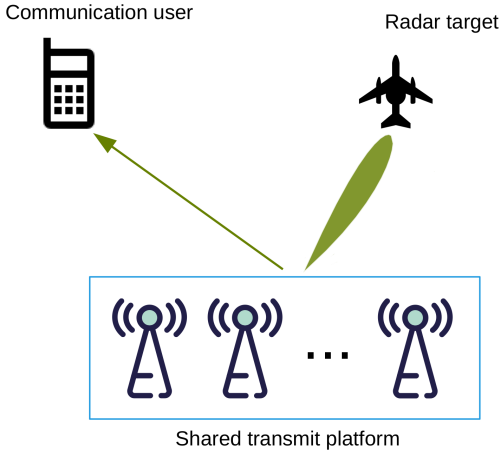}
	\caption{System model of a MIMO DFRC system \cite{lx2}}
	\label{fig:1}
\end{figure}

A \ac{MIMO} \ac{DFRC} system simultaneously performs \ac{MIMO} communication and \ac{MIMO} radar functions, whose waveform is optimized to meet requirements of both radar and communication \cite{mdy}. Denote by $M$ the number of transmit antennas, and by $\bm{x} \left( t \right) \in \mathbb{C}^{M}$ the transmit waveform. As illustrated in Fig.~\ref{fig:1}, to facilitate the theoretical analysis, in this paper we consider a toy example where there is only one communication user with a single receive antenna, one point-like radar target located at angle $\theta$ and no clutter in space. In addition, we assume that the transmit signal is narrow-band and the system uses the same transmit antennas to receive radar returns.

\subsection{Single-user \ac{MIMO} communication performance}

The received signal of the communication user $y_{C} \left( t \right) \in \mathbb{C}$ is expressed as
\begin{equation}
	y_{C} \left( t \right) =\bm{h}^{H}\bm{x} \left( t \right) +n_{C} \left( t \right),
\end{equation}
where $\bm{h} \in \mathbb{C}^{M}$ is the channel vector, and $n_C \left( t \right) \in \mathbb{C}$ is the \ac{AWGN} of the communication receiver. Without loss of generality, we normalize the power of the \ac{AWGN} by setting
$n_{C} \left( t \right) \sim \mathcal{CN} \left( 0,1 \right)$.

The achievable rate is a fundamental bound of communication performance. It is determined by the covariance matrix of the transmit signal, given by
\begin{equation}
	\bm{R}=\mathrm{E} \left\{ \bm{x} \left( t \right) \bm{x}^{H} \left( t \right) \right\}\in \mathbb{C}^{M \times M}.
	\label{eqR}
\end{equation}
The achievable rate from the transmitter to the communication user is calculated as \cite{capacity}
\begin{equation}
	C=\log \left( 1 +\frac{\bm{h}^{H}\bm{Rh}}{\sigma^{2}} \right) =\log \left( 1 +\bm{h}^{H}\bm{Rh} \right),
\end{equation}
where $\sigma^{2}$ is the power of the \ac{AWGN}, assumed to be 1.

\subsection{\ac{MIMO} radar performance}

Given the target direction $\theta$, the signal impinging on the target is $\bm{a}_{\bm{t}}^H \left( \theta  \right) \bm{x} \left( t \right)$, where $\bm{a_t} \left( \theta  \right) \in \mathbb{C}^{M}$ is the transmit steering vector. For a \ac{ULA}, $\bm{a_t}$ is given by
\begin{equation}
	\bm{a_t}\left(\theta\right)=\left[
	\begin{array}{cccc}
		1 & e^{j 2\pi \frac{d}{\lambda} \sin{\theta}} & \cdots & e^{j 2\pi \left(M-1\right) \frac{d}{\lambda} \sin{\theta}}
	\end{array}
	\right]^{H},
	\label{eqat}
\end{equation}
where $d$ is the interval between adjacent antennas and $\lambda$ is the wavelength. Then the received signal $\bm{y}_{R} \left( t \right) \in \mathbb{C}^{M}$ is expressed as
\begin{equation}
	\bm{y}_{R} \left( t \right) =\alpha_{0}\bm{a_{r}} \left( \theta  \right) \bm{a}_{\bm{t}}^H \left( \theta  \right) \bm{x} \left( t \right) +\bm{n}_{R} \left( t \right),
\end{equation}
where $\alpha_0$ is the target amplitude, $\bm{a_r} \left( \theta  \right) \in \mathbb{C}^{M}$ is the receive steering vector, 
$\bm{n}_{R} \left( t \right) \in \mathbb{C}^{M}$ is the \ac{AWGN} of the radar receiver, i.e. $\bm{n}_{R} \left( t \right) \sim \mathcal{CN} \left( 0,\bm{I}_M \right)$. We have $\bm{a_{r}} \left( \theta  \right) = \bm{a_{t}} \left( \theta  \right)$ if we use the same antenna array to receive radar returns. Here we also note that the Doppler frequency and time-delay of the target echo are omitted for simplicity. The received radar echo is beamformed to enhance the SNR, yielding
\begin{equation}
	\begin{aligned}
		r\left(t\right) & =\bm{w}^{H}\bm{y}_{R} \left( t \right) \\
		& =\alpha_{0}\bm{w}^{H}\bm{a_{r}} \left( \theta  \right) \bm{a}_{\bm{t}}^H \left( \theta  \right) \bm{x} \left( t \right) +\bm{w}^{H}\bm{n}_{R} \left( t \right),
	\end{aligned}	
\end{equation}
where $\bm{w} \in \mathbb{C}^{M}$ is the receive beamforming vector.

We use \ac{SNR} to evaluate the radar performance, which is given by \cite{snr}
\begin{equation}
	\begin{aligned}
		{\rm SNR} & = \mathrm{E} \left\{ \frac{\left| \alpha_{0}\bm{w}^{H}\bm{a_{r}} \left( \theta  \right) \bm{a}_{\bm{t}}^H \left( \theta  \right) \bm{x} \left( t \right) \right|^2}{\left| \bm{w}^{H}\bm{n}_{R} \left( t \right) \right|^2} \right\} \\
		& = \frac{\alpha_0^2\bm{w}^H\bm{a_{r}a}_{\bm{t}}^H\bm{Ra_{t}a}_{\bm{r}}^H\bm{w}}{\left\Vert \bm{w} \right\Vert_2 ^2}.
	\end{aligned}
	\label{eq2-5}
\end{equation}
Under the constraint $\left \Vert \bm w \right \Vert_2^2=1$, the maximal \ac{SNR} is achieved when $\bm{w}=\frac{\bm{a_r}}{\left\Vert \bm{a_r} \right\Vert_2}$. Plugging it into \eqref{eq2-5} we have
\begin{equation}
	{\rm SNR}=\alpha_0^2\left\Vert\bm{a_r}\right\Vert_2^2\bm{a}_{\bm{t}}^H\bm{Ra_{t}}.
\end{equation}

\subsection{Transmit power constraint}

We consider sum-power constraint
\begin{equation}
	\mathrm{E} \left\{ \left \Vert \bm{x} \left( t \right) \right \Vert_{2}^{2} \right\} \leq P,
	\label{eq2-1}
\end{equation}
where $P$ is the maximal average transmit power. By substituting \eqref{eqR} into \eqref{eq2-1}, we rewrite the power constraint with respect to $\bm{R}$ as
\begin{equation}
	\mathrm{tr} \left( \bm{R} \right) \leq P,
\end{equation}

\subsection{Problem formulation}

We formulate the fundamental performance bound of a \ac{MIMO} \ac{DFRC} system as an optimization problem. To this end, we calculate the theoretical performance bound of one function while constraining the other. Particularly, in this paper, we calculate the communication capacity under the \ac{SNR} constraint of the radar function. This yields the optimization problem
\begin{subequations}
	\begin{align}
		\max_{\bm{R}\succeq 0} \ & \ \log \left( 1 +\bm{h}^{H}\bm{Rh} \right), \label{eq2-6a} \\
		\mathrm{s.t.} \ & \ \alpha_0^2\left\Vert\bm{a_r}\right\Vert_2^2\bm{a}_{\bm{t}}^H\bm{Ra_{t}} \geq {\rm SNR_0}, \label{eq2-6b} \\
		\ & \ \mathrm{tr} \left( \bm{R} \right) \leq P, \label{eq2-6c}
	\end{align}
	\label{eq2-6}
\end{subequations}
where $\rm SNR_0$ is the allowed minimal radar \ac{SNR}.

Since the objective function \eqref{eq2-6a} monotonically increases with $\bm{h}^{H}\bm{Rh}$, which is the power of the signal transmitted to the communication user, maximizing \eqref{eq2-6a} is equivalent to maximizing $\bm{h}^{H}\bm{Rh}$. The radar \ac{SNR} is directly proportional to $\bm{a}_{\bm{t}}^H\bm{Ra_{t}}$, which is the power of the signal impinging on the radar target. Hence the constraint \eqref{eq2-6b} can be rewritten as $\bm{a}_{\bm{t}}^{H}\bm{Ra_{t}} \geq \frac{{\rm SNR_0}}{\alpha_0^2\left\Vert\bm{a_r}\right\Vert_2^2} = \gamma$, where $\gamma$ is the minimal power impinging on the radar target. In addition, since larger transmit power always leads to higher channel capacity, we let the sum-power constraint \eqref{eq2-6c} hold with equality. Therefore, optimization problem \eqref{eq2-6} is equivalently expressed as
\begin{subequations}
	\begin{align}
		\max_{\bm{R}\succeq 0} \ & \ \bm{h}^{H}\bm{Rh}, \label{eq2-7a}\\
		\mathrm{s.t.} \ & \ \bm{a}_{\bm{t}}^{H}\bm{Ra_{t}} \geq \gamma, \\
		\ & \ \mathrm{tr} \left( \bm{R} \right) = P.
	\end{align}
	\label{eq2-7}
\end{subequations}

\section{Beamforming design and performance analysis}\label{s3}

By solving the optimization problem \eqref{eq2-7}, we obtain the channel capacity and the corresponding waveform to achieve it. In this section, we provide the detailed derivations, which yield analytical expressions of the solution, offering insight into the \ac{MIMO} \ac{DFRC} system.

\subsection{Optimal beamforming design}\label{s3-a}

Note that the optimization problem \eqref{eq2-7} is a \ac{SDP} problem with two linear constraints. According to \cite{rank}, there exists an optimal $\bm{R}$ with rank one. Thus, we rewrite $\bm{R}$ as
\begin{equation}
	\bm{R}=\bm{cc}^{H},
	\label{eq3-1}
\end{equation}
where $\bm{c} \in \mathbb{C}^M$. Substituting \eqref{eq3-1} into \eqref{eq2-7} yields
\begin{subequations}
	\begin{align}
		\max_{\bm{c}} \ & \ \left| \bm{c}^{H}\bm{h} \right|^2, \\
		\mathrm{s.t.} \ & \ \left| \bm{c}^{H}\bm{a_{t}} \right|^2 \geq \gamma, \label{eq3-2b} \\
		\ & \ \bm{c}^{H}\bm{c} = P. \label{eq3-2c}
	\end{align}
	\label{eq3-2}
\end{subequations}

Solving the \ac{KKT} conditions of \eqref{eq3-2}, we find that $\bm{c}$ is expressed as
\begin{equation}
	\bm{c}=a\bm{h}+b\bm{a_t},
\end{equation}
where $a$, $b \in \mathbb{C}$ are given by
\begin{itemize}
	\item [i)]
	When $0\leq\gamma<\frac{P\left|\bm{h}^{H}\bm{a_t}\right|^2}{\left\Vert\bm{h}\right\Vert_2^2}$,
	\begin{subequations}
		\begin{numcases}{}
			\left|a\right| = \frac{\sqrt{P}}{\left\Vert \bm{h} \right\Vert_2}, \\
			\left|b\right| = 0,
		\end{numcases}
		\label{sol-1}
	\end{subequations}
	where the phase of $a$ is arbitrary.
	
	\item[ii)]
	When $\frac{P\left|\bm{h}^H\bm{a_t}\right|^2}{\left\Vert \bm{h} \right\Vert_2^2}\leq \gamma \leq P\left\Vert \bm{a_t} \right\Vert_2^2$,
	\begin{subequations}
		\begin{numcases}{}
			|a| = \eta, \label{sol-2a} \\
			|b| = \frac{\sqrt{\gamma}}{\left\Vert\bm{a_t}\right\Vert_2^2}-\frac{\left|\bm{h}^H\bm{a_t}\right|}{\left\Vert\bm{a_t}\right\Vert_2^2} \eta, \label{sol-2b}
		\end{numcases}
		\label{sol-2}
	\end{subequations}  
	where 
	\begin{equation}
		\eta = \sqrt{\frac{P\left\Vert\bm{a_t}\right\Vert_2^2-\gamma}{\left\Vert\bm{h}\right\Vert_2^2\left\Vert\bm{a_t}\right\Vert_2^2-\left|\bm{h}^H\bm{a_t}\right|^2}},
	\end{equation}
	and the phases of $a$ and $b$ should satisfy
	\begin{equation}
		\arg\left(a\right)-\arg\left(b\right)=\arg\left(\bm{h}^H\bm{a_t}\right)
		\label{sol-2-2}
	\end{equation}
	if $\bm{h}^H\bm{a_t} \neq 0$ or can be arbitrary if $\bm{h}^H\bm{a_t} = 0$.
	
	\item[iii)]
	When $\gamma > P\left\Vert\bm{a_t}\right\Vert_2^2$, there is no feasible solution. 
\end{itemize}

The derivation is provided in Appendix \ref{a-a}.

The formation of $\bm{c}$ indicates that the design of the optimal waveform is actually power allocation between the radar target and the communication user. The coefficients $a$ and $b$ represent the amplitudes of resources allocated for communication and radar sensing respectively. In case (i), the requirement for the radar sensing \ac{SNR} is low and $\left|b\right| = 0$. In this case, we do not need to particularly allocate power to the radar target, as the signal transmitted to the communication user already sheds enough power on the radar target to meet the \ac{SNR} requirement. In case (ii), the requirement for the radar sensing \ac{SNR} is higher. In this case, as the \ac{SNR} requirement increases, $\gamma$ increases, thereby $\left|a\right|$ decreases and $\left|b\right|$ increases, which means that we need to allocate more power to radar sensing. When $\gamma = P\left\Vert \bm{a_t} \right\Vert_2^2$, $\left|a\right| = 0$, the multi-antenna transmitter works like a phased-array radar and the radar \ac{SNR} achieves its upper bound. In case (iii), there is no feasible solution, since we can only achieve limited radar \ac{SNR} with limited transmit power.

When the radar target and the communication user are in the same direction, which means that $\bm{a_t}$ is parallel to $\bm{h}$, the communication waveform also serves as the radar sensing waveform, and there is no need to allocate the transmit power. When $\bm{a_t}$ is orthogonal to $\bm{h}$, radar sensing and communication functions operate independently without any energy sharing between each other. The expressions of the power allocated for communication and radar sensing are $\left| a \right|^2 \left\Vert \bm{h} \right\Vert_2^2$ and $\left| b \right|^2 \left\Vert \bm{a_t} \right\Vert_2^2$ respectively, which satisfy the equation $\left| a \right|^2 \left\Vert \bm{h} \right\Vert_2^2 + \left| b \right|^2 \left\Vert \bm{a_t} \right\Vert_2^2 = P$.

\subsection{Channel capacity analysis}

By substituting the optimal solution $\bm{R}$ into \eqref{eq2-6a}, we obtain the channel capacity with respect to radar \ac{SNR} threshold $\gamma$ as follows:
\begin{itemize}
	\item [i)] When $0\leq\gamma<\frac{P\left|\bm{h}^{H}\bm{a_t}\right|^2}{\left\Vert\bm{h}\right\Vert_2^2}$,
	\begin{equation}
		C=\log \left(1+P\left\Vert \bm{h} \right\Vert_2^2\right).
		\label{eq3-27a}
	\end{equation}
	The channel capacity is constant.
	\item[ii)] When $\frac{P\left|\bm{h}^H\bm{a_t}\right|^2}{\left\Vert \bm{h} \right\Vert_2^2}\leq \gamma \leq P\left\Vert \bm{a_t} \right\Vert_2^2$,
	\begin{equation}
		C=\log \left(1+\frac{\left( \sqrt{\gamma}\left| \bm{h}^{H}\bm{a_t} \right| + \sqrt{\beta} \right)^2}{\left\Vert\bm{a_t}\right\Vert_2^4}\right),
		\label{eq3-27b}
	\end{equation}
	where
	\begin{equation}
		\beta=\left( P\left\Vert \bm{a_t} \right\Vert_2^2-\gamma \right)\left( \left\Vert \bm{h} \right\Vert_2^2 \left\Vert \bm{a_t} \right\Vert_2^2 - \left|\bm{h}^H\bm{a_t} \right|^2 \right).
		\label{eq3-28}
	\end{equation}
	\item[iii)] When $\gamma > P\left\Vert\bm{a_t}\right\Vert_2^2$, the transmit power is not enough to meet the radar \ac{SNR} requirement.
\end{itemize}

In case (i), since there is no dedicated radar signal (the communication signal simultaneously serves as a probing function), the channel capacity is constant and equals the achievable capacity of the pure \ac{MIMO} communication system without radar function. In case (ii), we see that $\frac{\partial C}{\partial \gamma} \leq 0$, which means the channel capacity is negatively correlated with the threshold $\gamma$, which indicates the performance tradeoff between radar sensing and communication.

\section{Numerical results}\label{s4}

In this section, we demonstrate the performance of MIMO DFRC systems via numerical simulations. In the simulations, we consider a \ac{ULA} transmitter with half-wavelength element spacing $d=\frac{\lambda}{2}$, and let the number of the transmit antennas $M=10$. The radar target is located at angle $\theta=-30^{\circ}$. The communication signal is transmitted directly from the transmitter to the communication user. Given the communication user direction $\theta_C$, the channel vector $\bm{h}$ is given by
\begin{equation}
	\bm{h}=\left[
	\begin{array}{cccc}
		1 & e^{j\pi \sin{\theta_C}} & \cdots & e^{j 9 \pi \sin{\theta_C}}
	\end{array}
	\right]^{H}.
	\label{eq4-2}
\end{equation}
In the simulation, we use \ac{SNR} loss to represent the \ac{SNR} threshold in \eqref{eq2-6}, given by 
\begin{equation}
	{\rm SNR\enspace loss}=10\log_{10} \left( \frac{{\rm SNR_0}}{{\rm SNR_{MAX}}} \right),
	\label{eq6-7}
\end{equation}
where ${\rm SNR_{MAX}} = \alpha_0^2\left\Vert\bm{a_r}\right\Vert_2^{2}P\left\Vert \bm{a_t} \right\Vert_2^2$ is the maximal achievable SNR under the sum-power constraint.

\begin{figure}
	\centering
	\includegraphics[width=0.8\linewidth]{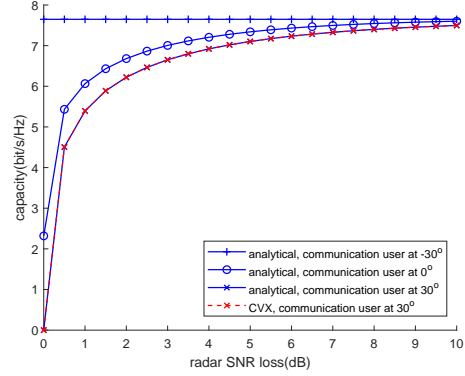}
	\caption{Channel capacity versus radar \ac{SNR} loss}
	\label{fig:2}
\end{figure}

We perform the first simulation to show the channel capacity versus radar \ac{SNR} loss with $\theta_C$ set to $-30^{\circ}$, $0^{\circ}$ and $30^{\circ}$. The results are shown in Fig.~\ref{fig:2}.  Since both \eqref{eq2-6} and \eqref{eq2-7} are convex and can be solved by the CVX, a package for specifying and solving convex programs \cite{cvx,gb08}, we also compare the the analytical expressions of the optimal solutions given in Section \ref{s3}  with the CVX results, denoted by `analytical' and `CVX', respectively. We find that for the same $\theta_C$, the CVX and analytical expressions yield the same curve, verifying the correctness of the derivation in Section~\ref{s3}.

From Fig.~\ref{fig:2}, the channel capacity increases with the radar \ac{SNR} loss when the communication user and the radar target are in different directions, $\theta \neq \theta_C$, while the capacity is fixed when $\theta = \theta_C$. We explain the phenomenon under different locations of down-link user and target. 
\begin{itemize}
    \item When $\theta=-30^{\circ}$ and $\theta_C=30^{\circ}$, it holds that $\bm{h}^{H}\bm{a_t}=0$. In this case, there is no energy sharing between radar and communication functions. As a result, the channel capacity reduces to 0 when the radar \ac{SNR} loss is $0$, because all the transmit energy is allocated to the radar. As the radar \ac{SNR} loss becomes higher, the channel capacity increases and reaches its upper bound, the channel capacity of a communication-only system with  all the transmit power  allocated to communication.
    \item When $\theta = \theta_C$, which means that $\bm{a_t}$ is parallel to $\bm{h}$, the transmit power is shared between these two functions. Consequently, by allocating all the transmit power in the desired direction, the channel capacity is fixed and achieves the upper bound, and the radar \ac{SNR} is also maximized simultaneously.
    \item When $\theta_C = 0^{\circ}$, the steering vector $\bm{a_t}$ is neither parallel nor perpendicular to $\bm{h}$. Part of the transmit power for communication is also used for probing. Therefore, even when the \ac{SNR} loss is strictly 0, the capacity is still positive. The capacity curve stays between those of $\theta_C = -30^{\circ}$ and $30^{\circ}$. 
\end{itemize}

\begin{figure}
	\centering
	\subfloat[Radar target at $-30^{\circ}$, communication user at $-30^{\circ}$\label{fig:3a}]
	{\includegraphics[width=0.8\linewidth]{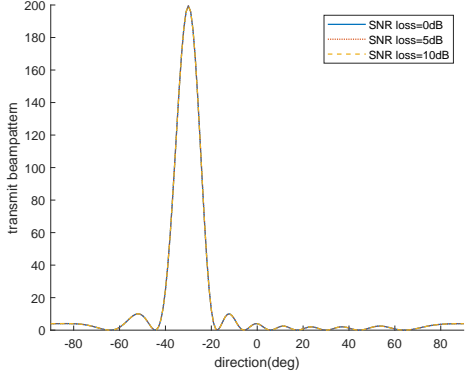}}
	\hfil
	\subfloat[Radar target at $-30^{\circ}$, communication user at $0^{\circ}$\label{fig:3b}]
	{\includegraphics[width=0.8\linewidth]{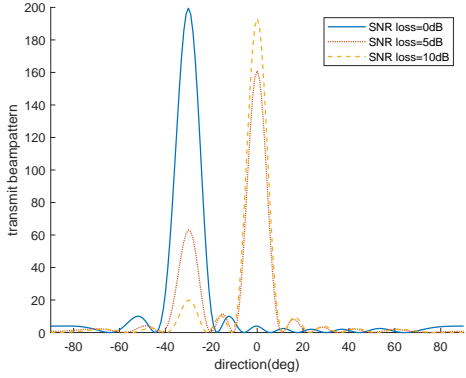}}
	\caption{Transmit beam pattern for different \ac{SNR} loss}
	\label{fig:3}
\end{figure}

In the second simulation, we show the transmit beam patterns for different radar \ac{SNR} losses in Fig.~\ref{fig:3a} and \ref{fig:3b}, where $\theta_C$ are set to $-30^{\circ}$ and $0^{\circ}$, respectively. In Fig.~\ref{fig:3a}, when the communication user and  radar target are in the same direction, $\theta = \theta_C$, the transmit beam pattern is  towards $\theta$ for different radar \ac{SNR} losses. In this case, there is no power allocation between radar and communication, because the communication signal simultaneously serves for  probing. In Fig.~\ref{fig:3b}, when $\theta \neq \theta_C$, there are two beams towards $\theta$ and $\theta_C$, respectively, indicating that the transmit power is allocated between radar and communication functions. As the radar \ac{SNR} loss increases, less power is allocated to the radar target.

We note that our results are different from those in  \cite{tb}, which states that there is \ac{SNR} loss for radar when the communication user and the radar target are in the same direction. The reason is that in this paper we use capacity to evaluate the communication performance, not restricting to using a certain coding strategy for communication. However, the signalling scheme in \cite{tb} constrains that the down-link user receives desired signals, which causes \ac{SNR} loss for radar. 


\section{Conclusion}\label{s5}

In this paper, we consider a toy example of a \ac{MIMO} \ac{DFRC} system where there is only one down-link user with a single receive antenna and one radar target. We take achievable communication rate as the communication performance metric, and radar \ac{SNR} as the radar performance metric. By constraining radar performance and optimizing communication performance, we formulate an optimization problem to study the performance bound of the system. We find the analytical expressions of the optimal waveform design and corresponding channel capacity, which facilitate the discussion on the performance bound of \ac{MIMO} \ac{DFRC} systems. The \ac{DFRC} system essentially allocates transmit power between radar target and down-link users, revealing the performance tradeoff between these two functions.

\appendices

\section{Solving the optimization \eqref{eq3-2}}\label{a-a}

Before solving \eqref{eq3-2}, we analyze the feasibility of the optimization problem first. According to \eqref{eq3-2c}, we have
\begin{equation}
	\bm{a}_{\bm{t}}^H\bm{cc}^{H}\bm{a_{t}} \leq \left\Vert \bm{c} \right\Vert_2^2 \left\Vert \bm{a_t} \right\Vert_2^2 = P\left\Vert \bm{a_t} \right\Vert_2^2,
\end{equation}
where the equivalence is achieved when $\bm{c}$ is parallel to $\bm{a_t}$. Thus the $\gamma$ in \eqref{eq3-2b} must satisfy that
\begin{equation}
	\gamma \leq P\left\Vert \bm{a_t} \right\Vert_2^2.
\end{equation}
Otherwise there is no feasible solution for \eqref{eq3-2}, which proves the result of case (iii) in Section \ref{s3-a}.

Since \eqref{eq3-2} satisfies the Slater's condition, we solve it by considering its \ac{KKT} conditions

The Lagrange function of \eqref{eq3-2} is given by
\begin{equation}
	\begin{aligned}
		& L \left( \bm{c},\bm{c}^{*},\lambda,\mu \right) \\ & =-\bm{h}^{H}\bm{cc}^H\bm{h}+\lambda \left( \gamma-\bm{a}_{\bm{t}}^H\bm{cc}^H\bm{a_{t}} \right) +\mu \left( \mathrm{tr} \left( \bm{cc}^H \right) -P \right),
	\end{aligned}
\end{equation}
where $\lambda$ and $\mu$ are dual variables. Then its \ac{KKT} conditions are
\begin{subequations}
	\begin{numcases}{}
		\frac{\partial L\left(\bm{c},\bm{c}^{*},\lambda,\mu\right)}{\partial \bm{c}^{*}}=-\bm{h}^H\bm{ch}-\lambda\bm{a}_{\bm{t}}^H\bm{ca_t}+\mu\bm{c}=\bm{0}, \label{eq3-4a}\\
		\mathrm{tr}\left(\bm{cc}^H\right)-P=0, \label{eq3-4b}\\
		\gamma-\bm{a}_{\bm{t}}^H\bm{cc}^H\bm{a_{t}}\leq0, \label{eq3-4c}\\
		\lambda\geq0, \label{eq3-4d}\\
		\lambda\left(\gamma-\bm{a}_{\bm{t}}^H\bm{cc}^H\bm{a_{t}}\right)=0. \label{eq3-4e}
	\end{numcases}
	\label{eq3-4}
\end{subequations}

To solve \eqref{eq3-4}, we first analyse the form of the optimal $\bm{c}$. Rewrite \eqref{eq3-4a} as
\begin{equation}
	\bm{c}=\frac{\bm{h}^H\bm{c}}{\mu}\bm{h}+\frac{\lambda\bm{a}_{\bm{t}}^H\bm{c}}{\mu}\bm{a_t},
	\label{eq3-5}
\end{equation}
indicating that $\bm{c}$ is a linear combination of $\bm{h}$ and $\bm{a_t}$. Assume that
\begin{equation}
	\bm{c}=a\bm{h}+b\bm{a_t}.
	\label{eq3-6}
\end{equation}
We then calculate $a$ and $b$ by substituting \eqref{eq3-6} into \eqref{eq3-4}.

According to \eqref{eq3-4e}, at least one of $\lambda$ and $\gamma-\bm{a}_{\bm{t}}^H\bm{cc}^H\bm{a_{t}}$ equals $0$. Consider the case that $\lambda=0$. Then \eqref{eq3-5} becomes
\begin{equation}
	\bm{c}=\frac{\bm{h}^H\bm{c}}{\mu}\bm{h}.
	\label{eq3-19}
\end{equation}
Substituting \eqref{eq3-6} into \eqref{eq3-4b} and \eqref{eq3-19} yields
\begin{subequations}
	\begin{numcases}{}
		\left|a\right|^2\bm{h}^H\bm{h}+a^*b\bm{h}^H\bm{a_t}+ab^*\bm{a}_{\bm{t}}^{H}\bm{h}+\left|b\right|^2\bm{a}_{\bm{t}}^{H}\bm{a_t}=P, \label{eq3-20a}\\
		a\bm{h}+b\bm{a_t}=\frac{a\bm{h}^H\bm{h}+b\bm{h}^H\bm{a_t}}{\mu}\bm{h}. \label{eq3-20b}
	\end{numcases}
	\label{eq3-20}
\end{subequations}
\noindent According to \eqref{eq3-20b} we find that
\begin{equation}
	b=0.
	\label{eq3-21}
\end{equation}
By substituting \eqref{eq3-21} into \eqref{eq3-20a} we obtain that
\begin{equation}
	\left|a\right|=\frac{\sqrt{P}}{\left\Vert \bm{h} \right\Vert_2},
	\label{eq3-23}
\end{equation}
where the phase of $a$ is arbitrary. Substitute the solution of $\bm{c}$ into \eqref{eq3-4c}. We then find that
\begin{equation}
	\gamma \leq \frac{P\left|\bm{h}^H\bm{a_t}\right|^2}{\left\Vert \bm{h} \right\Vert_2^2},
	\label{eq3-25}
\end{equation}
which proves the result of case (i) in Section \ref{s3-a}.

Next, consider the case that 
\begin{equation}
	\gamma-\bm{a}_{\bm{t}}^H\bm{cc}^H\bm{a_{t}}=0.
	\label{eq3-27}
\end{equation}
By substituting \eqref{eq3-6}, we rewrite \eqref{eq3-27} and \eqref{eq3-5} as
\begin{subequations}
	\begin{numcases}{}
		\begin{aligned}
			& \left|a\right|^2\bm{a}_{\bm{t}}^H\bm{hh}^H\bm{a_t}+a^*b\bm{a}_{\bm{t}}^H\bm{a_{t}h}^H\bm{a_t} \\
			& \quad+ab^*\bm{a}_{\bm{t}}^{H}\bm{ha}_{\bm{t}}^H\bm{a_t}+\left|b\right|^2\bm{a}_{\bm{t}}^{H}\bm{a_t}\bm{a}_{\bm{t}}^{H}\bm{a_t}=\gamma,
		\end{aligned} \label{eq3-7b}
		\\
		\begin{aligned}
			& a\bm{h}+b\bm{a_t} \\
			& \quad=\frac{a\bm{h}^H\bm{h}+b\bm{h}^H\bm{a_t}}{\mu}\bm{h}+\frac{\lambda\left(a\bm{a}_{\bm{t}}^H\bm{h}+b\bm{a}_{\bm{t}}^H\bm{a_t}\right)}{\mu}\bm{a_t}.
		\end{aligned} \label{eq3-7c}
	\end{numcases}
	\label{eq3-7}
\end{subequations}
Combining \eqref{eq3-20a} and \eqref{eq3-7b} we obtain that
\begin{equation}
	\left|a\right|=\eta:=\sqrt{\frac{P\left\Vert \bm{a_t} \right\Vert_2^2-\gamma}{\left\Vert \bm{h} \right\Vert_2^2\left\Vert \bm{a_t} \right\Vert_2^2-\left|\bm{h}^H\bm{a_t} \right|^2}}.
	\label{eq3-11}
\end{equation}
From \eqref{eq3-7c} we find that
\begin{subequations}
	\begin{numcases}{}
		\arg \left(a\right) = \arg \left(a\bm{h}^H\bm{h}+b\bm{h}^H\bm{a_t}\right), \label{eq3-11-a} \\
		\arg \left(b\right) = \arg \left(a\bm{a}_{\bm{t}}^H\bm{h}+b\bm{a}_{\bm{t}}^H\bm{a_t}\right). \label{eq3-11-b}
	\end{numcases}
	\label{eq3-11-}
\end{subequations}
\noindent This infers the the relationship between the phases of $a$ and $b$. When $\bm{h}^H\bm{a_t}=0$, \eqref{eq3-11-} always holds and the phases of $a$ and $b$ are arbitrary. When $\bm{h}^H\bm{a_t} \neq 0$, we have
\begin{equation}
	\arg\left(a\right)-\arg\left(b\right)=\arg\left(\bm{h}^H\bm{a_t}\right).
	\label{eq3-12}
\end{equation}
This results from the facts that $\bm{h}^{H}\bm{h}$ in \eqref{eq3-11-a} is a real number and $\arg \left(a\right) = \arg \left(a\bm{h}^H\bm{h}\right)$, which implies $\arg \left(a\right) = \arg \left(b\bm{h}^H\bm{a_t}\right)$. Similarly, $\arg \left(b\right) = \arg \left(a\bm{a}_{\bm{t}}^H\bm{h}\right)$.


Substituting \eqref{eq3-11} and the relationship between the phases of $a$ and $b$ into \eqref{eq3-20a} yields
\begin{equation}
	\left|b\right|=\frac{\sqrt{\gamma}}{\left\Vert \bm{a_t} \right\Vert_2^2}-\frac{\left|\bm{h}^H\bm{a_t}\right|}{\left\Vert \bm{a_t} \right\Vert_2^2}\eta.
	\label{eq3-15}
\end{equation}

In addition, since $\left|b\right|\geq 0$, we have $\gamma \geq \frac{P\left|\bm{h}^H\bm{a_t}\right|^2}{\left\Vert \bm{h} \right\Vert_2^2}$. From \eqref{eq3-11}, \eqref{eq3-12} and \eqref{eq3-15}, we prove the result of case (ii) in Section \ref{s3-a}.

\bibliographystyle{IEEEtran}
\IEEEtriggeratref{2}
\bibliography{sample}

\end{document}